\documentclass{nature}
\usepackage{float}
\usepackage{color}
\usepackage{graphicx}
\usepackage{amsmath,caption}
\usepackage{amssymb} 
\usepackage[backend=bibtex, style=nature, natbib=true, sorting=none]{biblatex}
\usepackage{subfigure} 
\usepackage{epsfig}

\newcommand{\beq}{\begin{equation}}
\newcommand{\bea}{\begin{array}}
\newcommand{\beqa}{\begin{eqnarray}}
\newcommand{\eeq}{\end{equation}}
\newcommand{\eea}{\end{array}}
\newcommand{\eeqa}{\end{eqnarray}}
\newcommand{\nin}{\noindent}

\newcommand{\bfr}{\mbox{\boldmath $r$}}

\newcommand{\cendot}{\mbox{\boldmath $\cdot$}}

\title{Migration into a Companion's Trap:\\Disruption of Multiplanet Systems in Binaries}
\author{Jihad R. Touma$^1$ and  S. Sridhar$^2$}
\begin{document}
\maketitle
\begin{affiliations}
\item Department of Physics, American University of Beirut,
PO Box 11--0236, Riad El-Solh, Beirut 1107 2020, Lebanon
\item Raman Research Institute, Sadashivanagar, Bangalore 560 080, India
\end{affiliations}

\maketitle 

\noindent
{\bf 
Most exoplanetary systems in binary stars are of S--type, and consist of one or more planets orbiting a primary star with a wide binary stellar companion. Gravitational forcing of a single planet by a sufficiently inclined binary orbit can induce large amplitude oscillations of the planet's eccentricity and inclination through the Kozai-Lidov (KL) instability \cite{kozai62, lidov62}. KL cycling was invoked to explain: the large eccentricities of planetary orbits \cite{htt97}; the family of close--in hot Jupiters \cite{wm03, ft07}; and the retrograde planetary orbits in eccentric binary systems \cite{ln11, kdm11}. However, several kinds of perturbations can quench the KL instability, by inducing fast periapse precessions which stabilize circular orbits of all inclinations \cite{htt97}: these could be a Jupiter--mass planet, a massive remnant disc or general relativistic precession. Indeed, mutual gravitational perturbations in multiplanet S--type systems can be strong enough to lend a certain dynamical rigidity to their orbital planes \cite{tkr08}.  Here we present a new and faster process that is driven by this very agent inhibiting KL cycling. Planetary perturbations enable secular oscillations of planetary eccentricities and inclinations, also called Laplace--Lagrange (LL) eigenmodes \cite{md99}. Interactions with a remnant disc of planetesimals can make planets migrate, causing a drift of LL mode periods which can bring one or more LL modes into resonance with binary orbital motion. The results can be dramatic, ranging from excitation of large eccentricities and mutual inclinations to total disruption. Not requiring special physical or initial conditions, binary resonant driving is generic and could have profoundly altered the architecture of many S--type multiplanet systems. It can also weaken the multiplanet occurrence rate in wide binaries, and affect planet formation in close binaries.}


\noindent
The fiducial system has two planets on initially coplanar orbits around a solar mass primary star: an interior $10 M_{\rm Jup}$ planet on a circular orbit with an initial semi-major axis of $a_{\rm in}^i=5~{\rm AU}$, and an exterior $10 M_{\oplus}$ planet with initial semi--major axis  $a_{\rm out}^i$ between $8$ and $11~{\rm AU}$ and eccentricity $e_{\rm out}^i =0.05$. The binary is also a solar mass star with semi--major axis $a_b > 100~{\rm AU}$ and corresponding period $T_b$ and angular frequency $n_b$. Planetary migration driven by scattering of planetesimals has a long and productive history in relation to solar system archeology \cite{malhotra93, gltm05, hm05}. It is a complex process, as discussed in the Supplementary Notes. Here we use it in a simple manner: the outer planet is allowed to migrate outward due to interactions with a planetesimal disc, with its semi--major axis having a prescribed form, with characteristic timescale $\tau$. For the solar system, there are plausible arguments  that the migration time $\tau >10^8~{\rm yr}$ \cite{gltm05}, with a lower bound $\tau > 10^7~{\rm yr}$ arguably needed to recover the properties of the Neptune Trojans \cite{lykawkaetal09}; we assume $2.5\times 10^7~{\rm yr} < \tau < 5\times10^8~{\rm yr}$. The physics of the problem appears clearest in a secular setting wherein the fast planetary (but not the binary) orbital motions are averaged over, turning a point mass planet into a shape and orientation changing Gaussian wire \cite{md99}. We present a numerical simulation with a state--of--the--art, $N$--wire algorithm \cite{ttk09}, and develop a mathematical model to understand the results.

In the $N$--wire experiment the binary orbit was coplanar and circular, with $a_b=1000$ AU, and a period $T_b=22.36~{\rm Kyr}$. The initial periods of the two LL eigenmodes are $3.53~{\rm Myr}$ (a slow mode determined mainly by the massive inner planet) and $12.25~{\rm Kyr}$ (a fast mode reflecting mainly the precession of the outer planet). Outward migration of the outer planet slows down the faster LL mode until its period approaches the binary period $T_b=22.36~{\rm Kyr}$.  Fig.1a shows that the eccentricity of the outer planet $e_{\rm out}\simeq 0.05$ until $a_{\rm out} \simeq 11.89~{\rm AU}$. Then it is captured into a resonance and $e_{\rm out}$ begins increasing, rising to $0.51$ when $a_{\rm out}\simeq 15~{\rm AU}$ at time $t\simeq 4055~T_b$. Capture is also apparent in the behavior of the resonant argument, ${\phi}_{\rm res} (t) = {\varpi}_{\rm out}(t) - n_b t$, where ${\varpi}_{\rm out}$ is the apsidal longitude. From Fig.1b we see that, after a period of circulation, ${\phi}_{\rm res}$ enters into libration at resonance passage, with libration maintained for the full duration of the simulation. More details of the capture process are given in Extended Data Figs.1(a,b). 

This capture phenomenon is closely related to the lunar evection resonance, that may have played a significant role in shaping the early history of the lunar orbit \cite{tw98, cs12}. However, the \emph{LL--mode evection resonance}  (LLER) is a new process, so we also present an analytical model, valid for arbitrary binary eccentricity, in the Supplementary Notes. For a circular binary orbit, LLER dynamics is governed by the normal form Hamiltonian of eqn(\ref{eq:res-ham}): 
\beq
H_{\rm nf} =\delta \left (\frac{\xi^2 + \eta^2}{2}\right ) - \alpha'  {\left (\frac{\xi^2 + \eta^2}{2}\right )}^{2} - \beta'  \left (\frac{\xi^2 -\eta^2}{2}\right)\,,
\nonumber
\eeq
where $\eta$ and $\xi$ are a canonically conjugate pair of LL mode variables, and the parameters $\delta$, $\alpha'$ and $\beta'$ are functions of the slowly migrating $a_{\rm out}$. The theoretical prediction for the location of the exact resonance is shown as the dashed red curve in Fig.1a: exact resonance is first met by a zero eccentricity planet around $a_{\rm out} \simeq 11.875~{\rm AU}$; the planet circulating at $e_{\rm out}\simeq 0.05$ encounters resonance a bit later (when $a_{\rm out} \simeq 11.89~{\rm AU}$), then gets engulfed by a growing and migrating nonlinearly bounded resonance region. Prediction follows simulation in the mean until $e_{\rm out}\simeq 0.2$ in our 4th order model; a higher--order expansion will improve the fit between model and simulation. The evolving topology of flows in the $(\eta,\xi)$ phase space, along with key structural features in and around resonance are discussed in the Supplementary Notes. 

Whereas encounter with the LLER is certain with migration, capture in it is  probabilistic, and depends on the strength of the resonance, the migration 
rate and the initial planetary eccentricity at which LLER is encountered. 
For $\tau \sim10^4\,T_b$ and initial $e_{\rm out} = 0.05$, the probability of capture in LLER exceeds one--half. At the assumed binary separation, capture becomes certain as the migration rate is slowed down by two orders of magnitude; more details are discussed in the Supplementary Notes. Capture is likely to improve in tighter S--type systems: for $a_b=200~{\rm AU}$ and $a_{\rm in} = 5~{\rm AU}$, evection is crossed at $a_{\rm out} \simeq 7.43~{\rm AU}$, with faster apse--precession and a stronger resonance, in the course of migration. The capture probability remains high for initial $e_{\rm out} = 0.05$, even at faster migration rates. Having described the broad secular skeleton of LLER, we note that the full problem is richer due to the interplay of planetary mean--motion resonances (PMMR). To study this, it is necessary to perform simulations that do not average over planetary mean--motions. Below we present two such $N$--body simulations with the open--source package MERCURY \cite{cham99,  hm05}.

The first MERCURY experiment is the unaveraged version of the $N$--wire simulation of Fig.1, and its results are displayed in Fig.2. Signs of PMMR are apparent in Fig.2a, in the jumps experienced by semi--major axis of the outer planet as it migrates. The system is captured in LLER with consequent growth of the eccentricity (Fig.2b), and libration of the resonant argument (Extended Data Fig.2a). What is remarkable though,  and distinct from the $N$--wire experiment, is the non--monotonic behavior of the mean eccentricity of the LLER--locked planet, leading to escape from LLER altogether, eventually settling at $e_{\rm out}\simeq 0.12$. Escape from capture is due to planetary mean motion resonances (PMMR), which enhance exchange of angular momentum.  Of the four PMMR located near $a_{\rm out}\simeq 12.56~{\rm AU}$ the strongest is the 4:1, with argument $\phi_{4:1} = 4 \lambda_{\rm out} - \lambda_{\rm in} - 3 \varpi_{\rm out}\,$; a short time segment of $\phi_{4:1}$ is shown in Extended Data Fig.2b.

In the second MERCURY experiment, the binary orbit had eccentricity $0.4$ 
and inclination $40^{\circ}$, which are modest values for wide--binaries.
Fig.3a shows a 3:1 PMMR exciting $e_{\rm out}$ to $0.1$ at 
$t\simeq 21~{\rm Myr}$, with capture in LLER at $t \simeq 72~{\rm Myr}$ when $a_{\rm out} \simeq 11.92~{\rm AU}$. As earlier, passage through the 4:1 resonance forces the outer planet out of LLER. This is followed by another excitation around $120~{\rm Myr}$, associated with passage through a 9:2 resonance; and then through a cluster of resonances around $14.63~{\rm AU}$. Meanwhile $a_{\rm out}$ grows with jumps at the PMMR (Extended Data Fig.3a), and $\phi_{\rm res}$ transits in and out of libration during LLER (Extended Data Fig.3b). \emph{Both $a_{\rm out}$ and $e_{\rm out}$  diffuse until the planet is ejected from the system altogether}. Ejection is not a necessary outcome, but  is often associated with PMMR when both $a_{\rm out}$ and $e_{\rm out}$ are large. In Fig.3b we follow the excitation of the mutual inclination to $12^{\circ}$, due to coupling within LLER--lock, of eccentricity and inclination by a vertical resonance, which is followed by another excitation to $14.7^{\circ}$.

LLER is a powerful and generic mechanism that can profoundly affect the architecture of multiplanet S--type binary systems. It can also come in different flavors. {\bf 1.}~Inward migration of the inner planet can occur through a runaway process \cite{mhht98}, whose slower migration rate has higher capture probability, particularly in tighter S--type systems 
with shorter LL periods. Inward migration may explain the largish eccentricities and inclinations in systems with super--Jupiter sized planets on sub--AU orbits. {\bf 2.}~LLER--induced disruption in moderately wide binaries ($a_b < 1000~{\rm AU}$) may be responsible for the recently reported dearth of multiplanet systems in binaries at such separations \cite{wang14b}. The extent to which LLER disrupts/suppresses planet formation when $a_b<20~{\rm AU}$ \cite{wang14a} needs to assessed within planet formation studies \cite{rafikov14}. {\bf 3.}~A multi--mass planetary system will have a broader spectrum of LL frequencies than a two--planet system. The richer LLER and stronger PMMR open more pathways for disruption, and could relieve an initial multiplanet system of all but one of its planets. 
\printbibliography[keyword=main]


\begin{addendum}
\item[Acknowledgments] We are grateful to Scott Tremaine and the Institute for Advanced Study for hosting us in the early stages of our collaboration.
\item[Author Contributions] J.R.T. and S.S. identified the process, developed and analyzed mathematical models, and wrote paper and supplements. J.R.T. performed and analyzed numerical experiments, producing figures in article and supplements. 
\item[Author Information] The authors declare no competing financial interests. Correspondence and requests for materials should be addressed
to  J.R.T. (jihad.touma@aub.edu.lb).
\end{addendum}
\clearpage

\newpage
\begin{figure}
\noindent
\text{Figure 1}\par\medskip
\centering
 \includegraphics[scale=0.75]{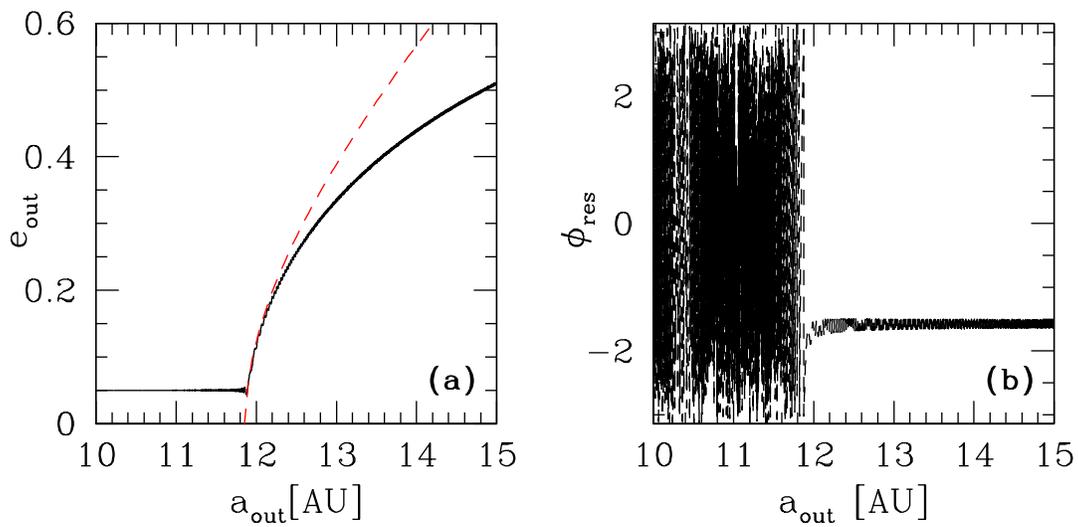}
\caption{{\bf Capture into the Laplace--Lagrange Evection Resonance.} The fiducial $N$--wire experiment was performed with exponential migration, $a_{\rm out}(t)  = a_{\rm out}^i \exp[t/\tau]$, with $a_{\rm out}^i=10~{\rm AU}$  and $\tau = 10^4 \,T_b$. {\bf (a)} Growth of $e_{\rm out}$ when it is captured in the migrating LLER. The dashed red line is the prediction from the analytical 4th--order theory presented in Supplementary Notes. {\bf (b)} $\phi_{\rm res}$ transitions from circulation to libration around $90^{\circ}$ when captured in LLER.}
\end{figure}

\newpage
\thispagestyle{empty}
\begin{figure}
\noindent
\text{Figure 2}\par\medskip
\centering
 \includegraphics[scale=0.75]{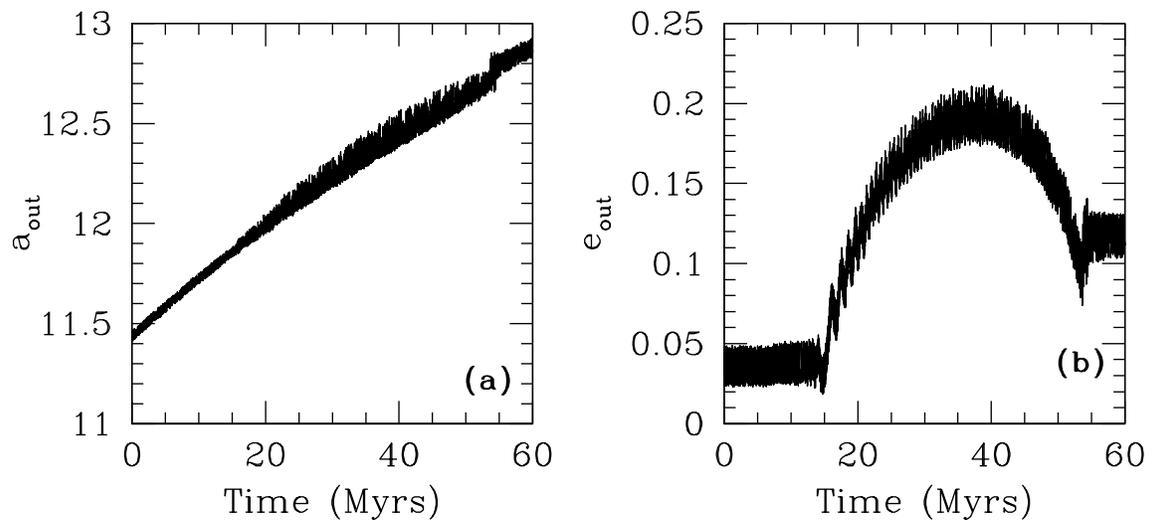}
\caption{{\bf LLER with PMMR for coplanar circular binary orbit.}
The first MERCURY experiment was conducted with damped migration, $a_{\rm out}(t)  = a_{\rm fin} - \Delta a\exp[-t/\tau]$, with $a_{\rm fin}=15~{\rm AU}$, 
$\Delta a = 3~{\rm AU}$ and $\tau = 10^8~{\rm yr}=4500\,T_b$. {\bf (a)} Migration of $a_{\rm out}$ with signs of PMMR. {\bf (b)} LLER is encountered around $a_{\rm out}\simeq 11.88~{\rm AU}\,$ at $\,t \simeq 15~{\rm Myr}$, with initial growth of $e_{\rm out}$ during capture, then decay because of interruption by PMMR, and ultimately escape from LLER.}
\end{figure}

\newpage
\thispagestyle{empty}

\begin{figure}
\noindent
\text{Figure 3}\par\medskip
\centering
\includegraphics[scale=0.75]{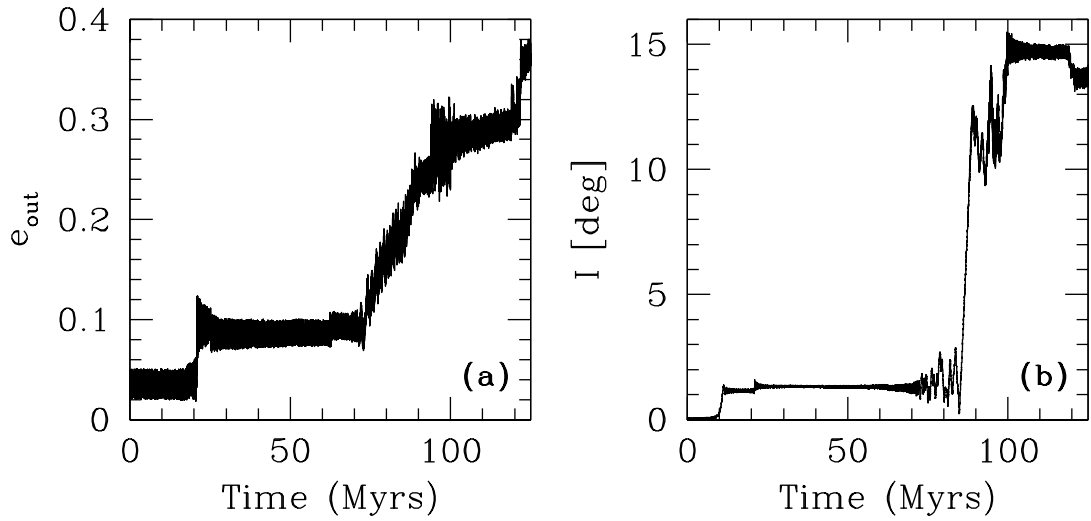}
\caption{{\bf LLER with PMMR for inclined and eccentric binary orbit.}
The second MERCURY experiment was conducted with exponential migration $a_{\rm out}(t) = a_{\rm out}^i \exp[t/\tau]$, with $a_{\rm out}^i= 10~{\rm AU}$ and $\tau = 4.4\times 10^8~{\rm yr} = 2\times 10^4 T_b$. {\bf (a)} The 3:1 PMMR around $21~{\rm Myr}$ excites $e_{\rm out}$ to about $0.1\,$. LLER is encountered around  $a_{\rm out}\simeq 11.92~{\rm AU}$, at $t\simeq 72~{\rm Myr}$.  {\bf (b)}  Mutual inclination first excited to $12^{\circ}$, then to $14.7^{\circ}$.}
\end{figure}
\clearpage
\newpage
\appendix
\setcounter{secnumdepth}{0}
\section{Supplementary Notes} 
\label{sec:modes}
\noindent
\centerline{\bf\large A Simple Analytical Model of the Laplace--Lagrange Evection Resonance}

\nin
We present an analytical model of the Laplace--Lagrange Evection Resonance
(LLER) in a two--planet system orbiting the primary star, with the companion star orbiting in the same plane as the planets. The main general result is
eqn.(\ref{ecchamfinal}) for the secular Hamiltonian for the two Laplace--Lagrange (LL) modes (given in the mode action--angle variables to
4th order in the planetary eccentricities) when they are forced by a binary orbit of arbitrary eccentricity. In order to understand the fiducial $N$--wire simulation reported in the main text we specialize to a circular binary orbit. Then the secular Hamiltonian reduces to the \emph{normal form}, $\,H_{\rm nf}$ of eqn.(\ref{eq:res-ham}). Simple computations with $H_{\rm nf}$ provide (a) a graphic narrative of the unfolding of LLER in phase space (Extended Data Fig.4); (b) characteristics of the LLER islands including measures of the adiabaticity (Extended Data Fig.5); (c) the dependence of capture probability on the 
initial planetary eccentricity and non--adiabaticity.

The Hamiltonian governing the secular dynamics of planets of mass $m_1$ and $m_2$ can be written as:  
\beqa
H_{\rm sec} &\;=\;& - Gm_1m_2\left<\!\!\!\left<\frac{1}{\left|\bfr_1 - \bfr_2\right|}\right>\!\!\!\right> \;-\; 
GM_B m_1\left<\frac{1}{\left|\bfr_b(t) - \bfr_1\right|}
- \frac{\bfr_b(t)\cendot\bfr_1}{r_b(t)^3}\right>
\nonumber\\[1ex]
&& -\; GM_B m_2\left<\frac{1}{\left|\bfr_b(t) - \bfr_2\right|}
- \frac{\bfr_b(t)\cendot\bfr_2}{r_b(t)^3}\right>\,.
\label{ham2sec}
\eeqa
Here ``$<\!\!<\,>\!\!>$'' means that the expression inside is to be averaged over the Kepler orbits of both planets, and ``$<\,>$'' means that the averaging is to be performed over the Kepler orbit of either planet 1 or 2, as the case may be.  Let $(a_1, a_2, a_b)\,$ and $(e_1, e_2, e_b)$ be the semi--major axes and the eccentricities of planet~1, planet~2 and the binary orbit, respectively. Let $g_1$ and $g_2$ be the periapse angles of the orbits of planets 1 and 2, and $\theta_b(t)$ be the polar angle to the location of the binary star. In the absence of planetary migration, secular dynamics conserves all the semi--major axes. $e_b$ is
constant, because the binary is assumed to be in a given Kepler orbit. The secular Hamiltonian governs the dynamics of the quantities $(g_1, g_2\,; e_1, e_2)\,$. We assume that $a_1 \leq a_2 \ll a_b$, and expand the binary potential to quadrupolar order: for $i=1,2$, 
\beq
\left<\frac{1}{\left|\bfr_b(t) - \bfr_i\right|}
- \frac{\bfr_b(t)\cendot\bfr_i}{r_b(t)^3}\right>
\;=\; 
\frac{a_i^2}{4r_b^3}\left[1 + \frac{3}{2}e_i^2 + 
\frac{15}{2}\cos{(2g_i - 2\theta_b)}\right] \;+\; \ldots\,.
\label{expr2}
\eeq 
However, a multipolar expansion of the interaction between planets 1
and 2 may not be appropriate, because their semi--major axes may be of comparable magnitudes. Therefore we assume that both $e_1\ll 1$ and $e_2\ll 1$ and expand to fourth order in the eccentricities: 
\beqa
\left<\!\!\!\left<\frac{1}{\left|\bfr_1 - \bfr_2\right|}\right>\!\!\!\right>
&\;=\;&
\frac{a_1}{2a_2^2} \Big\{c_{20}^{0}\,e_1^2  \;+\;   c_{11}^{1}\,e_1e_2\,
\cos{(g_1 - g_2)}  \;+\; c_{02}^{0}\,e_2^2 \;+\; c_{40}^{0}\,e_1^4  
\nonumber\\[1ex]
&& \quad  +\; c_{31}^{1}\,e_1^3 e_2\,\cos{(g_1 - g_2)} \;+\;
c_{22}^{0}\,e_1^2 e_2^2 \;+\; c_{22}^{2}\,e_1^2e_2^2\,\cos{(2g_1 - 2g_2)}
\nonumber\\[1ex]
&& \quad +\;   c_{13}^{1}\,e_1 e_2^3\,\cos{(g_1 - g_2)} \;+\; c_{04}^{0}\,e_2^4 \;+\; \ldots\Big\}\,,
\label{expr3}  
\eeqa 
where the $c_{nl}^{m}$ are functions of $(a_1/a_2)$, and can be written  
in terms of Laplace coefficients \cite{md99}. When  eqns.~(\ref{expr2}) 
and (\ref{expr3}) are substituted in eqn.~(\ref{ham2sec}), we obtain the secular Hamiltonian as a function of the four dynamical quantities $(g_1, g_2\,; e_1, e_2)\,$. However, these quantities are not canonically conjugate variables. Therefore we define new canonical coordinates $(q_1, q_2)$, and their canonically conjugate momenta $(p_1, p_2)$ by, 
\beqa
q_i &\;=\;& -\,\sqrt{2m_i\left(GM_A a_i\right)^{1/2}\left[1 - (1 - e_i^2)^{1/2}\right]\,}\;\sin{g_i}\,,\nonumber\\[1ex]
p_i &\;=\;& +\,\sqrt{2m_i\left(GM_A a_i\right)^{1/2}\left[1 - (1 - e_i^2)^{1/2}\right]\,}\;\cos{g_i}\,.
\eeqa
Expressing (\ref{expr2}) and (\ref{expr3}) in terms of the variables $(q_i, p_i)$, the secular Hamiltonian of eqn.~(\ref{ham2sec}) can be written as the sum of three terms:
\beq
H_{\rm sec} \;=\; H_{\rm LL} \;+\; H_{\rm bin} \;+\; H_{\rm non}\,.
\label{hsum3}
\eeq
Here $H_{\rm LL}$ is the Laplace--Lagrange Hamiltonian that consists of  all the time--independent quadratic terms, $H_{\rm bin}$ is the time--dependent driving due to the binary motion, and $H_{\rm non}$ is the nonlinear part that has all the time--independent fourth order terms.
\beq
H_{\rm LL} \;=\; \alpha_1\left(q_1^2 + p_1^2\right) \;+\; \beta\left(q_1q_2 + p_1p_2\right) \;+\; \alpha_2\left(q_2^2 + p_2^2\right)\,,
\label{hlldef}
\eeq
where the coefficients $\alpha_1$, $\beta$ and $\alpha_2$ are defined by,
\beqa
\alpha_1 &\;=\;& -\,\sqrt{\frac{G}{M_A}}\left[\frac{1}{2}\,\frac{m_2a_1^{1/2}c_{20}^{0}}{a_2^2} \;+\;  \frac{3}{8}\,\frac{M_Ba_1^{3/2}}{a_b^3\left(1-e_b^2\right)^{3/2}}\right]\,,
\nonumber\\[1ex]
\beta &\;=\;& -\,\frac{1}{2}\,\sqrt{\frac{Gm_1m_2}{M_A}}\,\frac{a_1^{3/4}c_{11}^{1}}{a_2^{9/4}}\,,\nonumber\\[1ex]
\alpha_2 &\;=\;& -\,\sqrt{\frac{G}{M_A}}\left[\frac{1}{2}\,\frac{m_1a_1c_{02}^{0}}{a_2^{5/2}} \;+\;  \frac{3}{8}\,\frac{M_Ba_2^{3/2}}{a_b^3\left(1-e_b^2\right)^{3/2}}\right]\,.
\label{constants1}
\eeqa
$H_{\rm LL}$ determines the two LL modes of oscillations of the eccentricities and periapses of the two planets. It has contributions from planetary interactions and the orbit--averaged binary quadrupole. To quadratic order the purely time--dependent binary forcing 
is represented by:
\beqa
H_{\rm bin} &\;=\;& \alpha_{b1}(t)\left[q_1^2 + p_1^2\right] \;+\; 
\gamma_{b1}(t)\left\{2q_1p_1\,\sin{\left[2\theta_b(t)\right]} + \left(q_1^2 - p_1^2\right)\,\cos{\left[2\theta_b(t)\right]}\right\} 
\nonumber\\
&+& \alpha_{b2}(t)\left[q_2^2 + p_2^2\right] \;+\; 
\gamma_{b2}(t)\left\{2q_2p_2\,\sin{\left[2\theta_b(t)\right]} + \left(q_2^2 - p_2^2\right)\,\cos{\left[2\theta_b(t)\right]}\right\}\,,
\label{hbindef}
\eeqa
where the constants $\alpha_{b1}(t)$, $\gamma_{b1}(t)$,
$\alpha_{b2}(t)$ and $\gamma_{b2}(t)$ are defined by, 
\beqa
\alpha_{b1}(t) &=& -\frac{3}{8}\sqrt{\frac{G}{M_A}}\,M_Ba_1^{3/2}\left[\frac{1}{r_b(t)^3} \;-\; \frac{1}{a_b^3\left(1-e_b^2\right)^{3/2}}\right]\,,\quad
\gamma_{b1}(t) = \frac{15}{8}\sqrt{\frac{G}{M_A}}\frac{M_Ba_1^{3/2}}{r_b(t)^3}\,,\nonumber\\[1ex]
\alpha_{b2}(t) &=& -\frac{3}{8}\sqrt{\frac{G}{M_A}}\,M_Ba_2^{3/2}\left[\frac{1}{r_b(t)^3} \;-\; \frac{1}{a_b^3\left(1-e_b^2\right)^{3/2}}\right]\,,\quad
\gamma_{b2}(t) = \frac{15}{8}\sqrt{\frac{G}{M_A}}\frac{M_Ba_2^{3/2}}{r_b(t)^3}\,.\nonumber\\[1ex]
\label{constants2}
\eeqa
When the binary orbit is circular, $r_b = a_b$ and $e_b = 0$; then the coefficients $\alpha_{b1}$ and $\alpha_{b2}$ both vanish. The nonlinear, 
time--independent,  fourth--order nonlinear terms are gathered together in:
\beqa
H_{\rm non} &\;=\;& \eta_1\left(q_1^2 + p_1^2\right)^2 \;+\;
\kappa_1\left(q_1q_2 + p_1p_2\right)\left(q_1^2 + p_1^2\right) \;+\;
\rho\left(q_1^2 + p_1^2\right)\left(q_2^2 + p_2^2\right) 
\nonumber\\
&+&\kappa_2\left(q_1q_2 + p_1p_2\right)\left(q_2^2 + p_2^2\right) \;+\;
\lambda\left(q_1q_2 + p_1p_2\right)^2 \;+\;
\eta_2\left(q_2^2 + p_2^2\right)^2\,,
\label{hnondef} 
\eeqa
where the constants $\eta_1$, $\kappa_1$, $\rho$, $\kappa_2$, $\lambda$ and 
$\eta_2$ are defined by, 
\beqa
\eta_1 &\;=\;& -\,\frac{1}{2}\,\frac{m_2}{M_Am_1a_2^2}\left(c_{40}^{0} \;-\; \frac{c_{20}^{0}}{4}\right)  \;+\; \frac{3}{32}\,\frac{M_Ba_1}{M_Am_1a_b^3\left(1-e_b^2\right)^{3/2}}\,,\nonumber\\[1ex]
\kappa_1 &\;=\;& -\,\frac{1}{2}\,\frac{m_2^{1/2}a_1^{1/4}}{M_Am_1^{1/2}a_2^{9/4}}\left(c_{31}^{1} \;-\; \frac{c_{11}^{1}}{8}\right)\,,\qquad\qquad
\rho \;=\; -\,\frac{1}{2}\,\frac{a_1^{1/2}}{M_Aa_2^{5/2}}\left(c_{22}^{0} \;-\; c_{22}^{2}\right)\,,\nonumber\\[1ex]
\kappa_2 &\;=\;& -\,\frac{1}{2}\,\frac{m_1^{1/2}a_1^{3/4}}{M_Am_2^{1/2}a_2^{11/4}}\left(c_{13}^{1} \;-\; \frac{c_{11}^{1}}{8}\right)\,,\qquad\qquad
\lambda \;=\; -\,\frac{a_1^{1/2}}{M_Aa_2^{5/2}}c_{22}^{2}\,,\nonumber\\[1ex]
\eta_2 &\;=\;& -\,\frac{1}{2}\,\frac{m_1a_1}{M_Am_2a_2^3}\left(c_{04}^{0} \;-\; \frac{c_{02}^{0}}{4}\right)  \;+\; \frac{3}{32}\,\frac{M_Ba_2}{M_Am_2a_b^3\left(1-e_b^2\right)^{3/2}}\,.
\label{constants3}
\eeqa
$H_{\rm non}$  determines the response of the LL modes to the resonant forcing by the binary.  The secular Hamiltonian, defined by eqns.~(\ref{hsum3})---(\ref{constants3}), governs the dynamics of the LL modes, in the 4--dimensional phase space, $(q_1, q_2\,; p_1, p_2)$.  Below we 
write each of its three terms, $H_{\rm LL}$, $H_{\rm bin}$ and 
$H_{\rm non}$, in terms of the action--angle variables of the LL modes.

\nin
{\bf 1a. Laplace--Lagrange modes:}
Define new canonical variables, $(Q_1, Q_2\,; P_1, P_2)$ by:\beqa
Q_1 &\;=\;& q_1\cos{\chi} + q_2\sin{\chi}\,,\qquad
Q_2 \;=\; -\,q_1\sin{\chi} + q_2\cos{\chi}\,;
\nonumber\\
P_1 &\;=\;& p_1\cos{\chi} + P_2\sin{\chi}\,,\qquad
P_2 \;=\; -p_1\sin{\chi} + p_2\cos{\chi}\,,
\label{cantr1}
\eeqa
where $\chi$ is such that $\;\tan{(2\chi)} \;=\; \beta/(\alpha_1 - 
\alpha_2)$\,. In the new variables, $H_{\rm LL}$ of eqn~(\ref{hlldef}) is:
\beqa
H_{\rm LL} &\;=\;& \left[\alpha_1\,\cos^2{\chi} + \beta\,\sin{\chi}\cos{\chi} + \alpha_2\,\sin^2{\chi}\right]\left(Q_1^2 + P_1^2\right)  \nonumber\\
&& +\; \left[\alpha_1\,\sin^2{\chi} - \beta\,\sin{\chi}\cos{\chi} + \alpha_2\,\cos^2{\chi}\right]\left(Q_2^2 + P_2^2\right)\,.
\label{hlltemp}
\eeqa
We now define action--angle variables for the LL modes, $(J_1, J_2\,; \psi_1, \psi_2)$, by:
\beq
Q_1 \;=\; \sqrt{2J}\,\sin{\psi_1}\,,\quad
P_1 \;=\; \sqrt{2J}\,\cos{\psi_1}\,,\quad
Q_2 \;=\; \sqrt{2J}\,\sin{\psi_2}\,,\quad
P_2 \;=\; \sqrt{2J}\,\cos{\psi_2}\,.
\label{llactang}
\eeq
Then  
\beq
H_{\rm LL} \;=\; \omega_1\,J_1 \;+\; \omega_2\,J_2\,,
\label{hllcanon}
\eeq
is in canonical form where 
\beqa
\omega_1 &\;=\;& (\alpha_1 + \alpha_2) \;+\; (\alpha_1 - \alpha_2)\cos{(2\chi)} \;+\; \beta\sin{(2\chi)}\,,\nonumber\\
\omega_2 &\;=\;& (\alpha_1 + \alpha_2) \;-\; (\alpha_1 - \alpha_2)\cos{(2\chi)} \;-\; \beta\sin{(2\chi)}\,,
\label{llfreqs}
\eeqa
are the mode frequencies. 

\nin
{\bf 1b. Resonant driving:} We use eqns.~(\ref{cantr1}) and (\ref{llactang}) to work out $H_{\rm bin}$ in terms of action--angle variables for the LL modes. Dropping the fourth order terms in eqn.~(\ref{hbindef}), we have
\beqa
H_{\rm bin} &\;=\;& \mu_1(t)\,J_1 \;+\; \mu_2(t)\,J_2 \;+\; \mu_3(t)\,\sqrt{J_1J_2}\,\cos{(\psi_1 - \psi_2)} \;+\; \nu_1(t)\,J_1\,\cos{\left[2\psi_1 + 2\theta_b(t)\right]} \nonumber\\
&& +\; \nu_2(t)\,J_2\,\cos{\left[2\psi_2 + 2\theta_b(t))\right]} \;+\; 
\nu_3(t)\,\sqrt{J_1J_2}\,\cos{\left[\psi_1 + \psi_2 + 2\theta_b(t)\right]}\,,
\label{hbinfinal}
\eeqa
where the new coefficients, $\mu_1(t)$, $\mu_2(t)$, $\mu_3(t)$, $\nu_1(t)$, 
$\nu_2(t)$ and $\nu_3(t)$, are defined by 
\beqa
\mu_1(t) &\;=\;& +\,2\alpha_{b1}(t)\,\cos^2{\chi} \;+\;  
2\alpha_{b2}(t)\,\sin^2{\chi}\,,\nonumber\\
\mu_2(t) &\;=\;& +\,2\alpha_{b1}(t)\,\sin^2{\chi} \;+\;  
2\alpha_{b2}(t)\,\cos^2{\chi}\,,\nonumber\\
\mu_3(t) &\;=\;& 2\left[\alpha_{b2}(t) \;-\; \alpha_{b1}(t)\right]\,\sin{(2\chi)}\,,\nonumber\\
\nu_1(t) &\;=\;& -\,2\gamma_{b1}(t)\,\cos^2{\chi} \;-\;  
2\gamma_{b2}(t)\,\sin^2{\chi}\,,\nonumber\\
\nu_2(t) &\;=\;& -\,2\gamma_{b1}(t)\,\sin^2{\chi} \;-\;  
2\gamma_{b2}(t)\,\cos^2{\chi}\,,\nonumber\\
\nu_3(t) &\;=\;& -\,2\left[\gamma_{b2}(t) \;-\; \gamma_{b1}(t)\right]\,\sin{(2\chi)}\,. 
\label{munudef}
\eeqa

\nin
{\bf 1c. Secular nonlinearities:}
Lastly, we write $H_{\rm non}$ in terms of LL--modal variables by using eqns.~(\ref{cantr1}) and (\ref{llactang}) to susbtitute for $\left(q_1, q_2\,; p_1, p_2\right)$ in terms of $\left(\psi_1, \psi_2\,; J_1, J_2\right)$ in eqn.~(\ref{hnondef}). Of the many terms, those proportional to 
$\cos{\left(\psi_1 - \psi_2\right)}\,$ and $\cos{\left(2\psi_1 - 2\psi_2\right)}\,$ can be dropped when $\omega_1$ and $\omega_2$ are well--separated (i.e. non--degenerate, as in the example explored in the body of the article), because the angle--dependent terms are oscillatory and do not contribute significantly to the dynamics. Therefore, nonlinear part of the Hamiltonian for nondegenerate modes can be taken as,
\beq 
H_{\rm non}^{\rm n.d.} \;=\; \xi_1\,J_1^2 \;+\; \xi_2\,J_2^2 \;+\; \xi_3\,J_1J_2\,, 
\label{hnonnd}
\eeq
where the coefficients, $\xi_1$, $\xi_2$ and $\xi_3$ are given by 
\beqa
\xi_1 &\;=\;& 4\eta_1\cos^4\chi \;+\; 2\kappa_1\cos^2\chi\,\sin(2\chi) \;+\;
\rho\sin^2(2\chi) \nonumber\\
&& +\; 2\kappa_2\sin^2\chi\sin(2\chi) \;+\; \lambda\sin^2(2\chi) \;+\; 4\eta_2\sin^4\chi\,,\nonumber\\[1ex]
\xi_2 &\;=\;& 4\eta_1\sin^4\chi \;-\; 2\kappa_1\sin^2\chi\,\sin(2\chi) \;+\;
\rho\sin^2(2\chi) \nonumber\\
&& -\; 2\kappa_2\cos^2\chi\sin(2\chi) \;+\; \lambda\sin^2(2\chi) \;+\; 4\eta_2\cos^4\chi\,,\nonumber\\[1ex]
\xi_3 &\;=\;& 4\eta_1\sin^2(2\chi) \;-\; 2\kappa_1\sin(4\chi) \;+\;
4\rho\cos^2(2\chi) \nonumber\\
&& +\; 2\kappa_2\sin(4\chi) \;+\; 2\lambda\cos(4\chi) \;+\; 
4\eta_2\sin^2(2\chi)\,.
\label{xicoeffsdef}
\eeqa

\nin
{\bf 2. Secular Hamiltonian for nondegenerate LL modes with binary driving:}
Gathering together with the expressions in eqns.~(\ref{hllcanon}), (\ref{hbinfinal}) and (\ref{hnonnd}), we have the secular Hamiltonian in the desired mode variables: 
\beqa
H_{\rm sec} &\;=\;& [\omega_1+ \mu_1(t)] \,J_1 \;+\; [\omega_2 + \mu_2(t)] \,J_2 \;+\; \xi_1\,J_1^2 \;+\; \xi_2\,J_2^2 \;+\; \xi_3\,J_1J_2 
\nonumber\\[1ex]
&& +\; \mu_3(t)\,\sqrt{J_1J_2}\,\cos{(\psi_1 - \psi_2)} \;+\; \nu_1(t)\,J_1\,\cos{\left[2\psi_1 + 2\theta_b(t)\right]} \;+ \nonumber\\[1ex]
&& +\; \nu_2(t)\,J_2\,\cos{\left[2\psi_2 + 2\theta_b(t))\right]} \;+\; 
\nu_3(t)\,\sqrt{J_1J_2}\,\cos{\left[\psi_1 + \psi_2 + 2\theta_b(t)\right]}\,.
\label{ecchamfinal}
\eeqa
There are resonances between the binary and the LL modes, when $n_b$ is commensurate with any of the frequencies $\omega_1$, $\omega_2$ or  $(\omega_1 + \omega_2)/2$. The set of resonances is particularly rich for an \emph{eccentric} binary orbit.  
When the binary orbit is \emph{circular}, the coefficients $\mu_i(t)$ all vanish, and the $\nu_i$ become time--independent. Then eqn.(\ref{ecchamfinal}) simplifies to:
\beqa
H_{\rm circ} &\;=\;& \omega_1\,J_1 \;+\; \omega_2\,J_2 \;+\; \xi_1\,J_1^2 \;+\; \xi_2\,J_2^2 \;+\; \xi_3\,J_1J_2 \;+\; \nu_{10}\,J_1\,\cos{\left(2\psi_1 + 2n_b t\right)}\nonumber\\[1ex]
&& \;\;+\;\;  \nu_{20}\,J_2\,\cos{\left(2\psi_2 + 2n_b t\right)} \;+\; 
\nu_{30}\,\sqrt{J_1J_2}\,\cos{\left(\psi_1 + \psi_2 + 2n_b t\right)}\,,
\label{sechamfinal}
\eeqa
where the coefficients, $\nu_{10}$, $\nu_{20}$ and $\nu_{30}$, are given by 
\beqa
\nu_{10} &\;=\;& -\,\frac{15}{4}\,\sqrt{\frac{G}{M_A}}\frac{M_B}{a_b^3}
\left(a_1^{3/2}\cos^2{\chi} \;+\; a_2^{3/2}\sin^2{\chi}\right)
\nonumber\\[1ex]
\nu_{20} &\;=\;& -\,\frac{15}{4}\,\sqrt{\frac{G}{M_A}}\frac{M_B}{a_b^3}
\left(a_1^{3/2}\sin^2{\chi} \;+\; a_2^{3/2}\cos^2{\chi}\right)
\nonumber\\[1ex]
\nu_{30} &\;=\;& -\,\frac{15}{4}\,\sqrt{\frac{G}{M_A}}\frac{M_B}{a_b^3}
\left(a_2^{3/2} \;-\; a_2^{3/2}\right)
\,\sin{(2\chi)}.
\label{nu0def}
\eeqa\emph{It turns out that $\omega_1$ and $\omega_2$ are negative, giving rise to three types of LLER, for $n_b\simeq\vert\omega_1\vert$, or $n_b\simeq\vert\omega_2\vert$, or $n_b\simeq\vert\left(\omega_1 + \omega_2\right)\!/2\vert\,$}. 
The possibilities are extremely rich, so we focus on the case relevant to the fiducial $N$--wire experiment described in the main text.

\nin
{\bf 3. Normal form Hamiltonian for the $\mbox{\boldmath $N$}\!$--wire experiment:}
In the $N$--wire experiment, the inner planet has $m_1 = 10~M_{\rm jup}$, $a_1 = 5$ AU with initial $e_1 = 0$, and $g_1=0$; the outer planet has $m_2 = 10~M_\oplus$, $a_2 < 
11.89~{\rm AU}$ (here it is initially at $11~{\rm AU}$) with $e_2= 0.05$ and $g_2 = 0$. The binary companion is also a solar mass star, on a circular orbit with semi-major axis $a_b = 1000~{\rm AU}$ and period $T_b =22,360.69~{\rm AU}$. The two LL mode frequencies are $\omega_{1} = -1.33\times10^{-6}$ rad/yr (a period of 4.73 Myrs) and $\omega_{2} = -3.95\times10^{-4}$ rad/yr (and a period of $15,918.29~{\rm yr}$). Here we study LLER when $n_b\simeq\vert\omega_2\vert$. Since  $\omega_1$ and $\omega_2$ are well--separated in magnitude, it is clear that $n_b$ cannot be close to either $\vert\omega_1\vert$ or $\vert\left(\omega_1 + \omega_2\right)\!/2\vert$. Then the driving terms proportional to $\cos{\left(2\psi_1 + 2n_b t\right)}$ and $\cos{\left(\psi_1 + \psi_2 + 2n_b t\right)}$ are oscillatory, and can be dropped. Hence $H_{\rm circ}$ is effectively independent of the angle $\psi_1$, which implies that $J_1 = J_{10} \simeq \mbox{constant}$. Therefore the resonant Hamiltonian for describing LLER of the second mode takes the simple form, 
\beq
H_{\rm m2} \;=\; \left(\omega_2 + n_b + \xi_3\,J_{10}\right)\,J_2 \;+\; \xi_2\,J_2^2  \;+\; \nu_{20}\,J_2\,\cos{\left(2\varphi_2\right)}\,.
\label{hammode2}
\eeq
In the absence of planetary migration, this is a time--independent 1 degree--of--freedom Hamiltonian in the canonically conjugate variables, $\varphi_2 = \psi_2 + n_b t$ and $J_2$, 
and  the dynamics is obviously integrable. This Hamiltonian is typical of 2nd--order resonance models, and can be further reduced to a \emph{normal form} in the new canonical variables, 
$\xi = \sqrt{2 J_2} \cos({\varphi}_{2})\,$ and $\,\eta = \sqrt{2 J_2} \sin(\varphi_2)$:
\beq
H_{\rm nf} =\delta \left (\frac{\xi^2 + \eta^2}{2}\right ) - \alpha'  {\left (\frac{\xi^2 + \eta^2}{2}\right )}^{2} - \beta'  \left (\frac{\xi^2 -\eta^2}{2}\right)\,,
\label{eq:res-ham}
\eeq
where
\beq
\delta \;=\;   \left(\omega_2 + n_b + \xi_3\,J_{10}\right)\,,\qquad
\alpha' \;=\; - {\xi}_{2}\,,\qquad
\beta'  \;=\; - \nu_{20}.
\label{alphetc}
\eeq
The normal form Hamiltonian has a long history in solar--system dynamics
(see \cite{bg84} and references therein). Of relevance to our problem is 
\emph{lunar evection}, the resonance between the precession of the peripase of the Moon's orbit around an oblate Earth, and the mean motion of a massive outer perturber, the Sun \cite{tw98}.  

\nin
{\bf 4. Planetary migration:}
Numerical simulations of planetary migration with gaseous discs have reported a wide variety of behaviour --- planets opening gaps, clearing out inner discs, stalling in their migration, reversing migration; multiple planets undergoing divergent migration, undergoing convergent migration, or getting captured into mean motion resonances \cite{aa09, armitage2011, hi13} --- but we do not explore this here. We consider \emph{planetary migration driven by scattering of planetesimals}. This process is believed to have taken place in the solar system, and to have left its signature in the dynamical properties of minor bodies, as well as spin and orbital features of the planets themselves \cite{malhotra93, fi84, hm99, gltm05, mltg05}. The planetary system is considered fresh out of the evaporation of the gaseous disc, with a remnant disc of surviving planetesimals which, in the course of their dynamical stirring, then scattering, by the planets is expected to drive migration. The timescale of migration is set by the inner boundary, mass and size distribution in the remnant disc \cite{hm99, gltm05}. For the solar system, there are plausible arguments for it being on the order of a few times $10^8$ years \cite{gltm05}, with a lower bound of a few~$10^7~{\rm yr}$ set by exercises which seek to recover properties of Neptune Trojans with planetary migration \cite{lykawkaetal09}. We cannot commit to any particular timescale without careful (and largely numerical) treatment of mean motion resonances and their stirring of a preexisting disc into planet crossing orbits. We have thus assumed a range of timescales $10^7~{\rm yr}$ to $4\times 10^8~{\rm yr}$, which is about $500$ to $20,000$ binary orbital periods. The direction of migration on the other hand is largely determined by the mass and location of the perturbing planets. 
 
\nin
{\bf 5. Dynamics of LLER:} As $a_2$ increases from an initial value of $11~{\rm AU}$ due to planetary migration, the control parameters $(\delta, \alpha', \beta')$
acquire slow time dependence, making  $H_{\rm nf}$ of eqn.(\ref{eq:res-ham}) a 1.5 degree--of--freedom system. In the $N$--wire fiducial system,   
$\delta$ is an increasing function of time, starting with a negative value  $-1.14\times 10^{-4}$ and then transitioning to positive values around $a_2= 11.884~{\rm AU}$; $\beta'\sim 10^{-6}$ is always positive; $\alpha'=2.24$  initially, and decreases while remaining positive over the relevant range of $a_2$. The variation of $(\delta, \alpha', \beta')$ results in non trivial changes in the topology of the instantaneous global phase portraits of $H_{\rm nf}$ in the $(\eta,\xi)$ plane. As can be seen in the four panels of 
Extended Data Fig.4, the origin $(0,0)$ --- corresponding to a circular orbit --- is always an equilibrium point. Since $\alpha' >0$, the origin is initially stable because $\delta < -\beta'$. 
As $\delta$ increases in the course of migration, it goes unstable for $\delta \geq -\beta'$, which happens at $a_2 = 11.875~{\rm AU}$.
This first bifurcation gives rise to two stable equilibria at $(\pm\eta_c, 0)$, where $\eta_c ={[(\delta+\beta')/\alpha']}^{1/2}$. These are the centres of LLER with librating orbits around them; see Figs.S1(a, b). Post--encounter, as $\delta$ continues to increase, the centres drift apart and the islands grow, capturing into LLER any trajectory that comes their way. Stability is restored to the origin for $\delta \geq +\beta'$ (at $a_2=11.89~{\rm AU}$). This second bifurcation gives rise to two unstable equilibria at $(0, \pm\xi_{\rm un})$ where $\xi_{\rm un} = {[(\delta-\beta')/\alpha']}^{1/2}$; see Extended Data Figs.4(c, d). As $\delta$ continues to grow, the basin of circulating orbits around the origin also grows, squeezing the LLER islands and capturing some of their librating orbits. Extended Data Fig.5a  shows the evolution of $\eta_c$, and the extrema of the separatrix. In Extended Data Figs.5(b,c) we map the evolution of $\eta_c$ to that of the planetary eccentricities. \emph{The dashed red curve in Fig.~1a of the main text is obtained from $e_2(t)$ of Extended Data Fig.5b}.

When $(\delta, \alpha', \beta')$ vary slowly compared to the libration period around LLER, we are in the \emph{adiabatic regime}. At any time, there 
is a maximum eccentricity, $e_{\rm max}(t)$, that is reached by the separatrix; let $e_c$ be the maximum value of all the $e_{\rm max}(t)$.  Capture is certain if LLER is encountered when $e_2 < e_c$.   In our problem, $e_c \simeq 0.054$, corresponding to the  the onset of the second bifurcation shown in Extended Data Fig.4c. If $e_2 > e_c$ at resonance encounter, capture is not certain. The probability of capture can be computed analytically \cite{bg84}, and is given by the ratio of (a) the rate of increase of the area of the libration zone, to (b) the rate of increase of the sum of the areas of the libration and circulation zones. Note that the circulation zone has zero area for $-\beta'/\alpha' \leq \delta/\alpha' \leq \beta'/\alpha'$, hence capture is certain with  $\delta/\alpha'$ increasing past $-\beta'/\alpha'$. When the variation is \emph{non--adiabatic}, outcomes are not easily predictable from the instantaneous phase portraits. Then capture and escape must be quantified through computations with $H_{\rm nf}$ for different initial conditions of the planet. 

\nin
{\bf 6. Estimates of Adiabaticity and Capture Probability:} 
We computed  $\Gamma = \omega_{\rm lib}/2\pi r_{\rm mig}$, where $r_{\rm mig} = {\rm d}\ln{\eta_c}/{\rm d}t$ is the migration rate of the island centres, and $\omega_{\rm lib} = 2\sqrt{\beta' \,[\delta+\beta']}$ is the libration frequency around the island centre. 
$\Gamma$ is a measure of adiabaticity, and is plotted versus $e_{\rm max}$ in Extended Data Fig.5.d. The dynamics is increasingly adiabatic for larger $a_2$, with the island migrating to higher eccentricities. For $\tau \sim10^4\,T_b$ and $e_{\rm max} = 0.054$, we have $\Gamma\simeq 0.2$; the migration rate is larger than the libration time, implying non--adiabatic passage (in the range of $e_{\rm max}$ for which capture is guaranteed in the adiabatic limit). Were $\tau$ larger by a factor $100$, we would be in the adiabatic regime, and capture in LLER would be certain for $e_2 \leq 0.054$ (excepting near--zero eccentricity where adiabaticity is practically impossible). Theoretical estimates of the capture probabilities are not well--determined in this non--adiabatic regime. Hence we integrated trajectories with the evolving $H_{\rm nf}$ for a range of initial eccentricities and uniformly distributed periapses, and discovered that: (a) Capture is ruled out for $e_2<0.028$; this outcome appears typical of non--adiabatic passage through LLER, and was already noted in studies of the early history of the lunar orbit \cite{tw98}; (b) Matters improve for larger eccentricities: more than half the planets with $e_2 =0.05$ get captured, and the capture probability rate gets closer to the adiabatic estimate with larger $e_2$ at encounter.
\thispagestyle{empty}
\printbibliography[keyword=supplement]

\newpage
\setcounter{figure}{0}
\renewcommand{\figurename}{Extended Data Figure}

\newpage
\thispagestyle{empty}
\begin{figure}
\noindent
\text{Extended Data Figure 1}\par\medskip
\centering
 \includegraphics[scale=0.75]{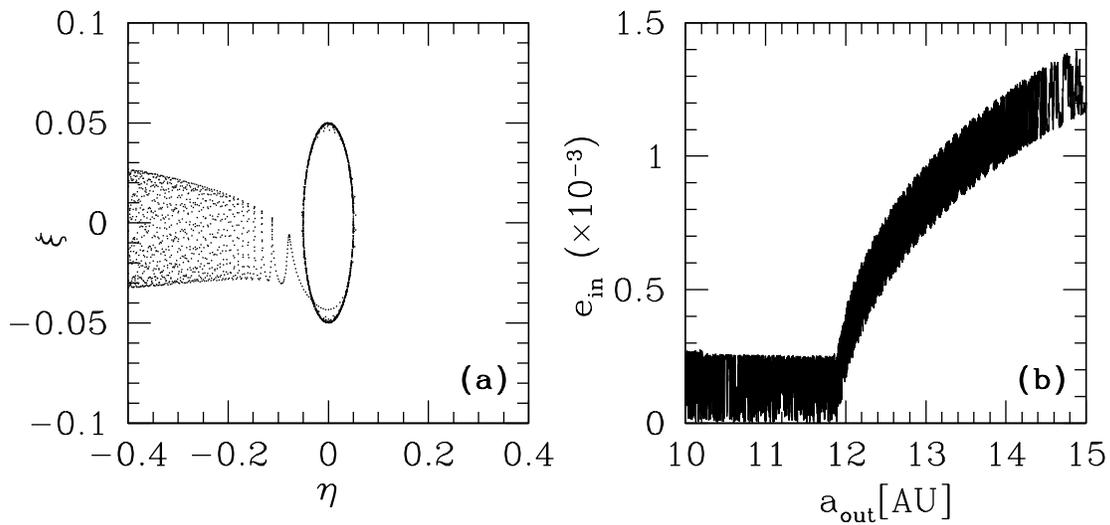}
\caption{{\bf Capture into the Laplace--Lagrange Evection Resonance.} The fiducial $N$--wire experiment was performed with exponential migration, $a_{\rm out}(t)  = a_{\rm out}^i \exp[t/\tau]$, with $a_{\rm out}^i=10~{\rm AU}$  and $\tau = 10^4 \,T_b$. {\bf (a)} Phase space trajectory of the outer planet during capture.   $\eta = [1 - \sqrt{1 - e_{\rm out}^2}] \sin(\phi_{\rm res})$ and $\xi=  [1 - \sqrt{1 - e_{\rm out}^2}] \cos ({\phi}_{\rm res}]$ are canonical coordinate and momentum of the captured LL mode. Trajectory clearly reveals transition from circulation with initial eccentricity of $0.05$ (inner ring) to libration when captured in LLER (funnel from inner ring moving toward negative $\eta$). {\bf (b)} Capture is also seen, albeit in a subdued manner, in the modest growth of $e_{\rm in}$ during LLER.}
\end{figure}
\clearpage
\newpage
\thispagestyle{empty}
\begin{figure}
\noindent
\text{Extended Data Figure 2}\par\medskip
\centering
 \includegraphics[scale=0.75]{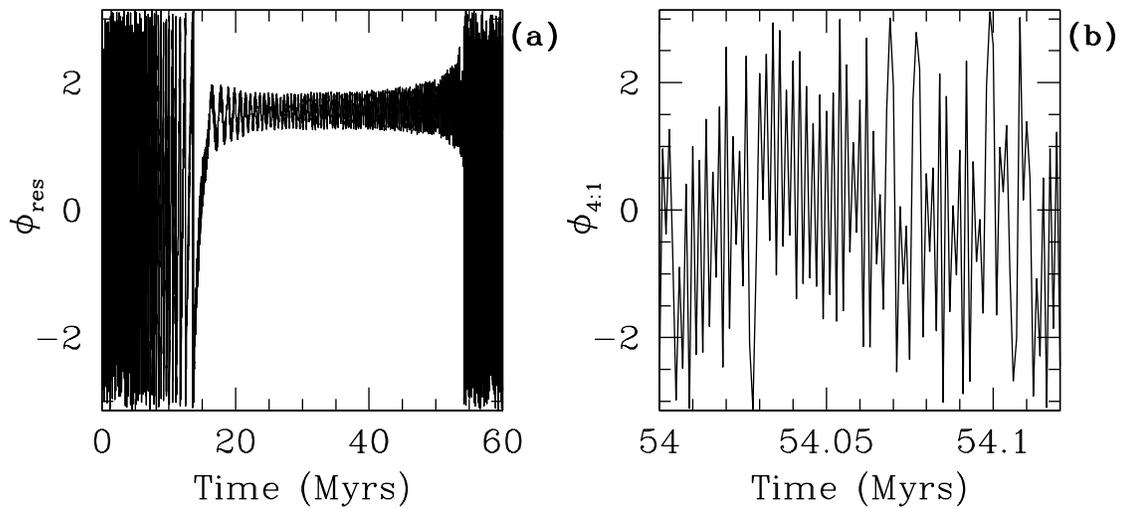}
\caption{{\bf LLER with PMMR for coplanar circular binary orbit.} The first MERCURY experiment was conducted with damped migration, $a_{\rm out}(t) = a_{\rm fin} - \Delta a\exp[-t/\tau]$, with $a_{\rm fin}=15~{\rm AU}$, $\,\Delta a = 3~{\rm AU}$ and $\tau = 10^8~{\rm yr}=4500\,T_b$. {\bf (a)} Transitions of $\phi_{\rm res}$ from circulation to libration during capture and back to circulation after escape. {\bf (b)} A $100~{\rm Kyr}$ time--segment of $\phi_{4:1}$ when still captured in evection; signature of the PMMR is evident in the repeated transitions  between libration and circulation.}
\end{figure}
\clearpage
\newpage
\thispagestyle{empty}
\begin{figure}
\noindent
\text{Extended Data Figure 3}\par\medskip
\centering
 \includegraphics[scale=0.75]{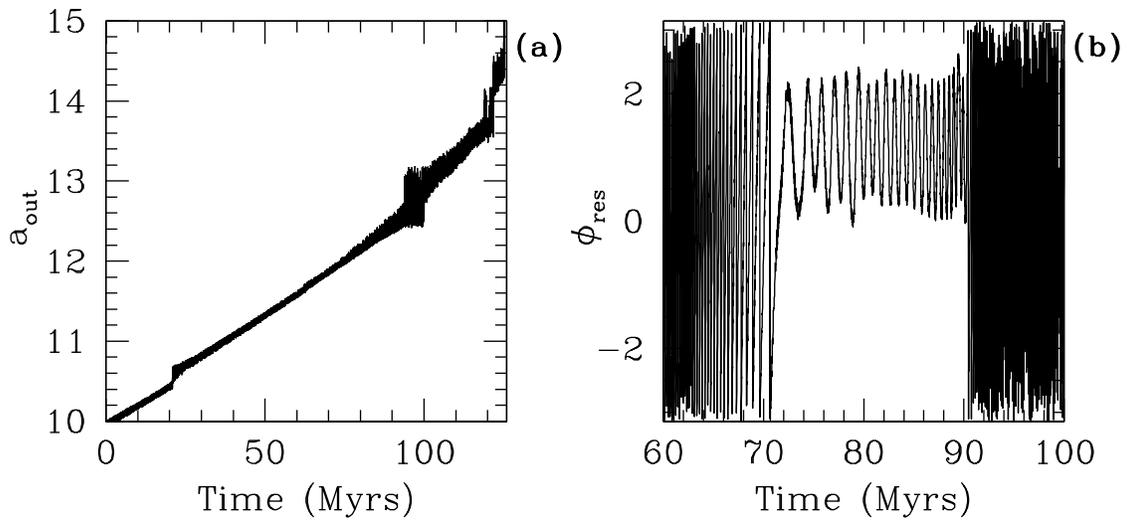}
\caption{{\bf LLER with PMMR for inclined and eccentric binary orbit.}The second MERCURY experiment was conducted with exponential migration $a_{\rm out}(t) = a_{\rm out}^i \exp[t/\tau]$, with $a_{\rm out}^i= 10~{\rm AU}$ and $\tau = 4.4\times 10^8~{\rm yr} = 2\times 10^4 T_b$. {\bf (a)} Migration of $a_{\rm out}$ with signs of PMMR around $21, 90, 120$ and $125~{\rm Myr}$. {\bf (b)} Transitions of $\phi_{\rm res}$ from circulation to libration during capture in LLER, and then back to circulation due to passage through the 4:1 PMMR which forces the outer planet out of LLER.}
\end{figure}
\clearpage
\newpage
\thispagestyle{empty}
\begin{figure}
\noindent
\text{Extended Data Figure 4}\par\medskip
\centering
\vspace{2cm}
\begin{minipage}{1\linewidth}
\centering
  \includegraphics[width=2.75in]{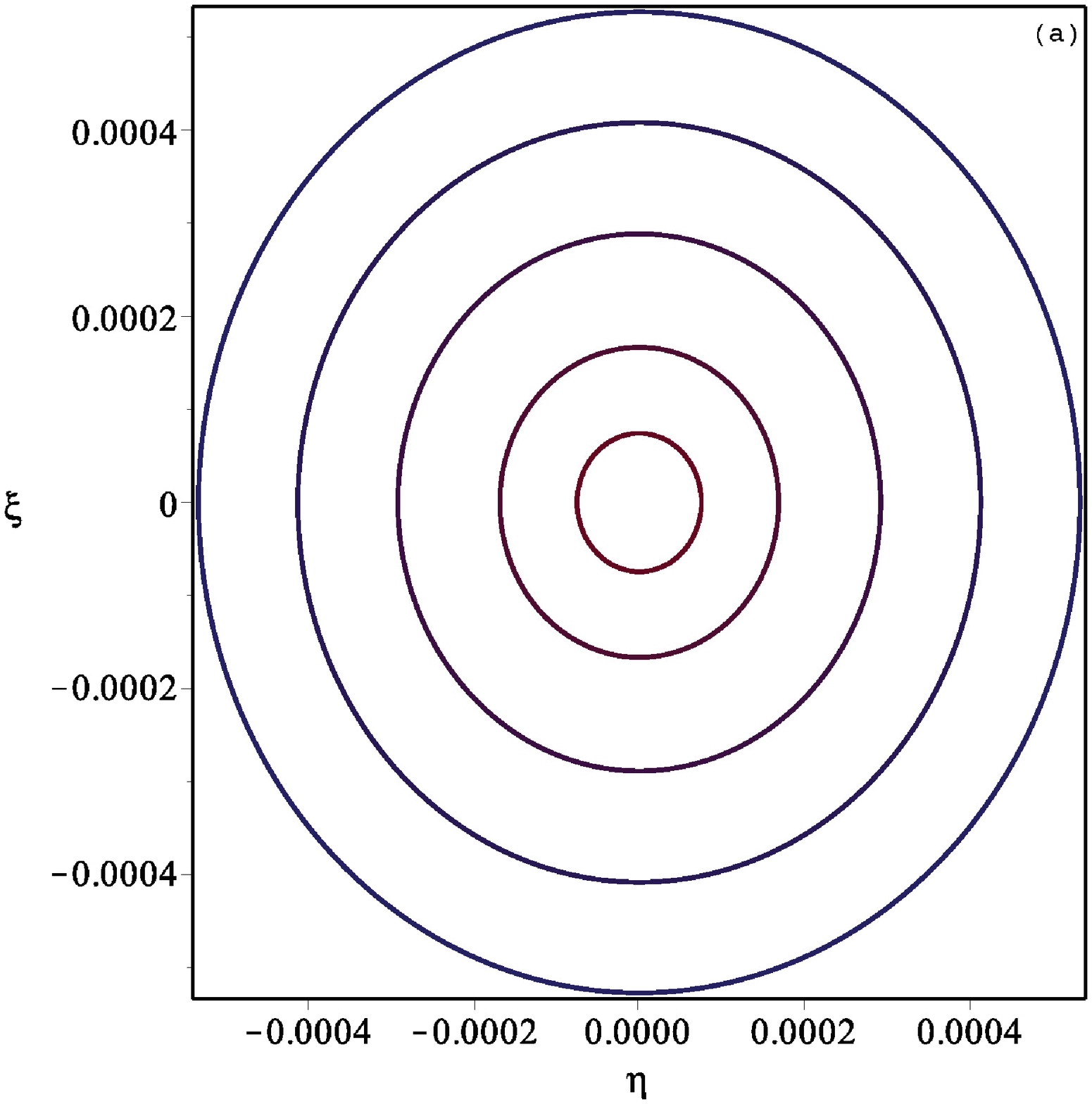} 
  \includegraphics[width=2.75in]{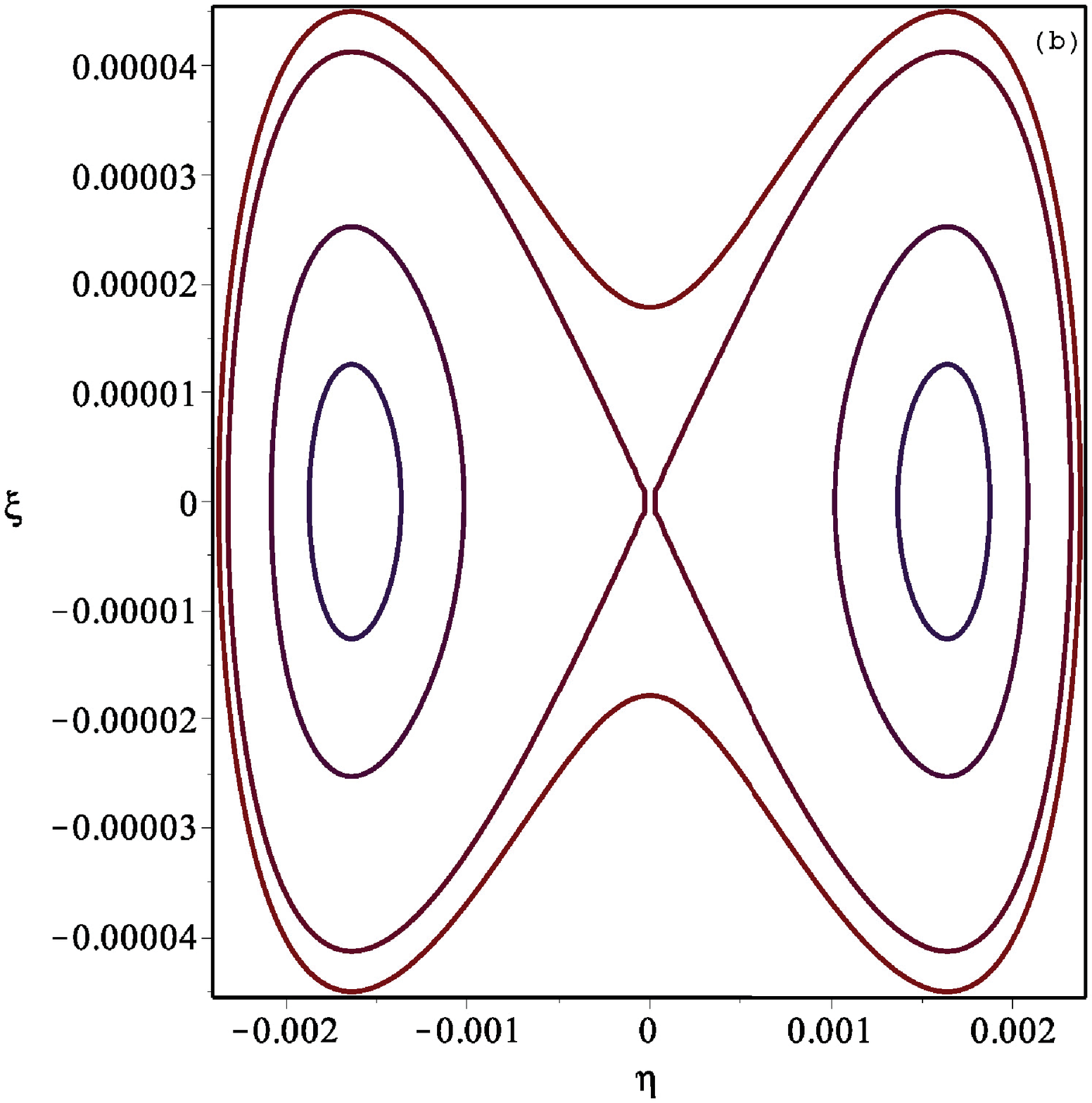} 
\end{minipage}
\begin{minipage}{1\linewidth}
\centering
 \includegraphics[width=2.75in]{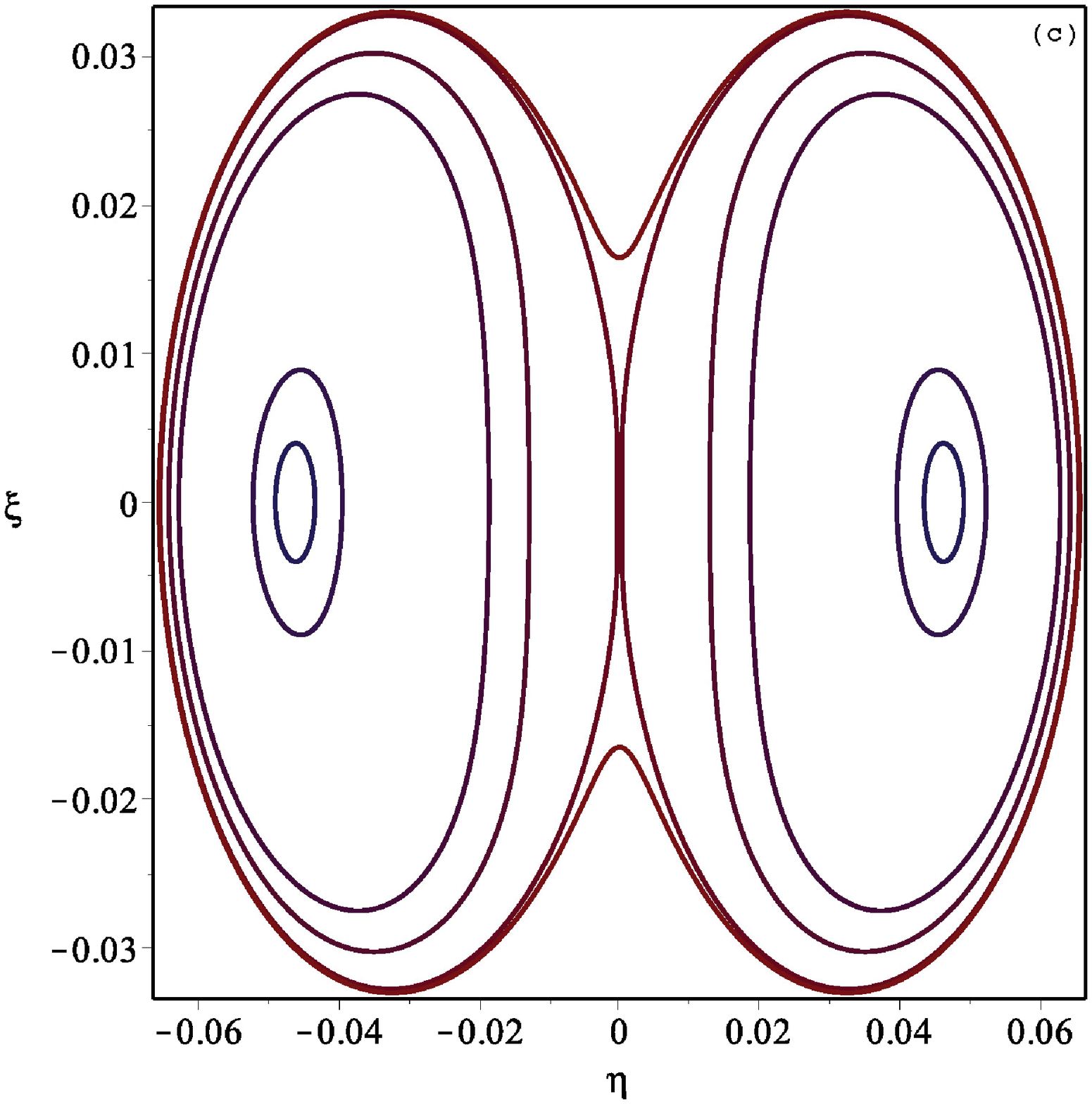}
 \includegraphics[width=2.75in]{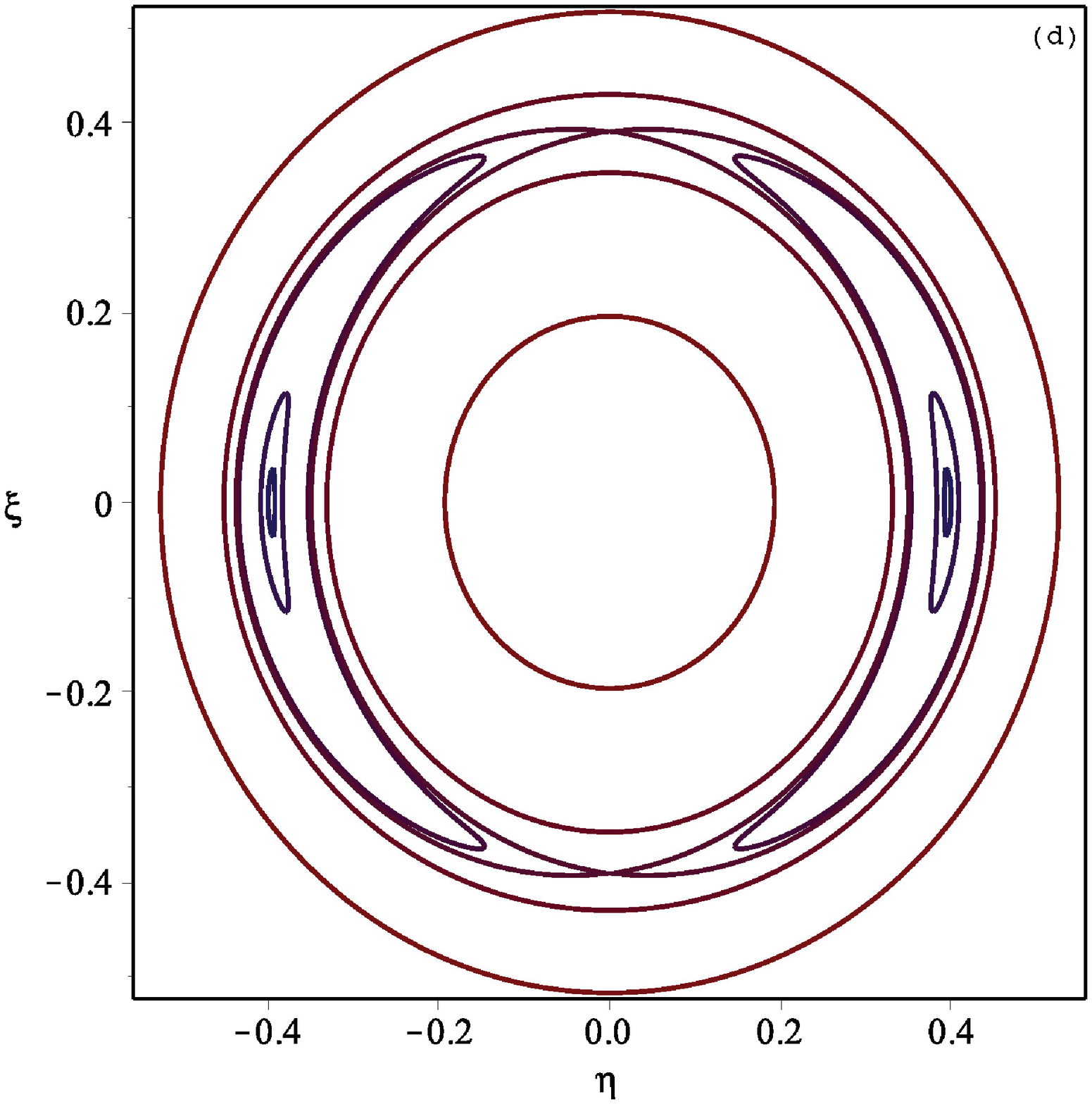}
\end{minipage}
\caption{{\bf Phase Space with Migrating Planet.} Isocontours of $H_{\rm nf}$ at different times, showing bifurcations of equilibria and emergence in islands where capture is probable. Note: both $\xi$ and $\eta$ have been  rescaled by a factor of $\sqrt{m_2\sqrt{G M_{A} a_2}}$ to turn them into eccentricity like variables. {\bf (a)} At $a_2=11~{\rm AU}$ the origin is stable with circulating orbits around it. {\bf (b)} At $a_2=11.88~{\rm AU}$ the origin has gone unstable, and two LLER islands have appeared.
{\bf (c)} At $a_2=11.894~{\rm AU}$ the origin is about to go stable again. 
{\bf (d)} At $a_2=13~{\rm AU}$ we are past the second bifurcation; there is an inner circulating zone surrounded by two libration lobes.}
\end{figure} 
\clearpage
\newpage
\thispagestyle{empty}
\begin{figure}
\noindent
\text{Extended Data Figure 5}\par\medskip
\vspace{2cm}
\centering
\includegraphics[scale=0.85]{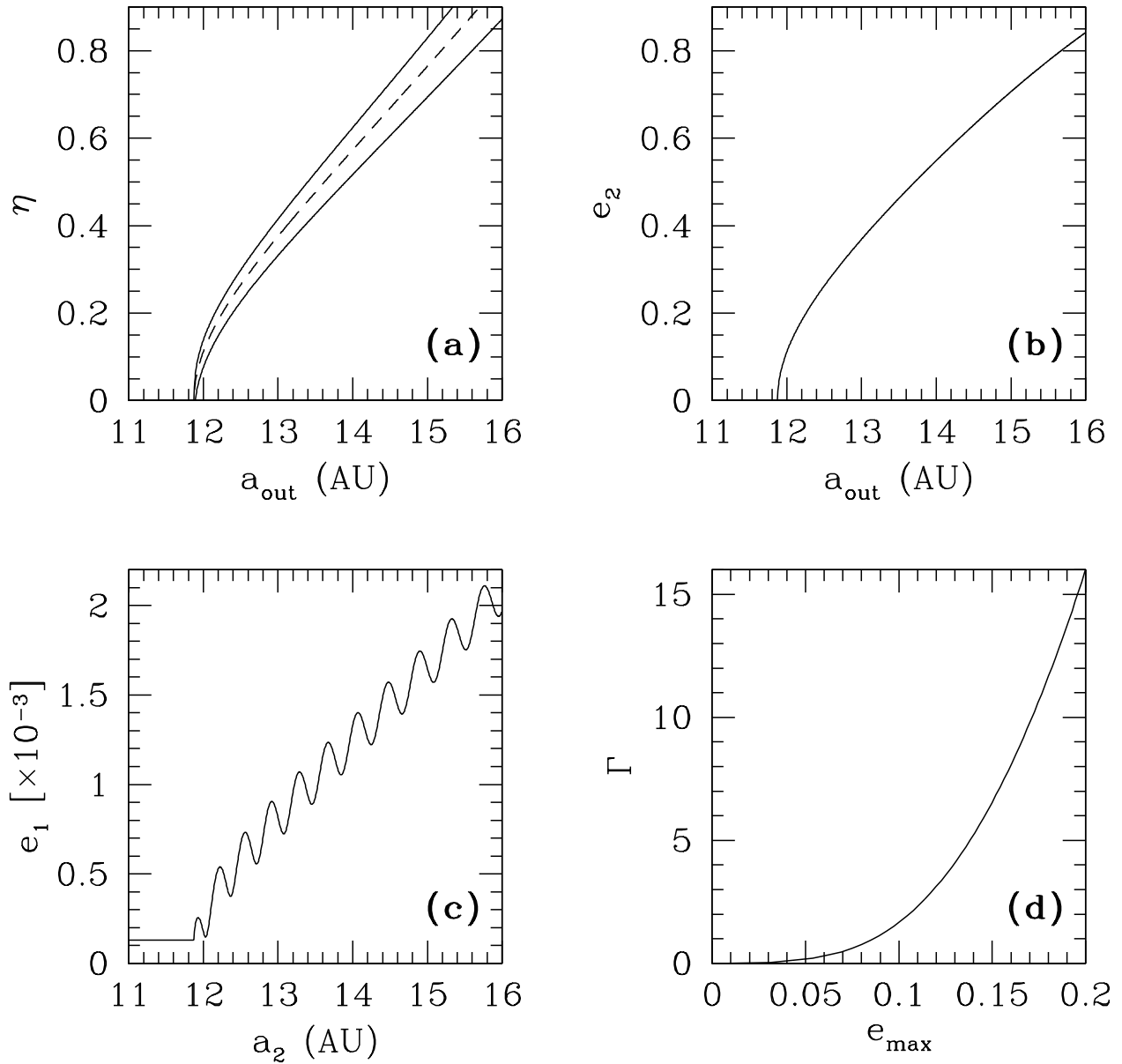}
\caption{ {\bf Characteristics of the LLER Islands:} 
{\bf (a)} Drift of LLER centers and extremities (again $\eta$ is rescaled by a factor of $\sqrt{m_2\sqrt{G M_{A} a_2}}$ to turn it into an eccentricity--like variable). 
{\bf (b)} The eccentricity of the outer planet increases significantly
during capture in LLER. This is the dashed red line in Fig.1 which is compared with the fiducial $N$--wire simulation.
{\bf (c)} Modest growth of the eccentricity of the inner planet when captured in LLER.
{\bf (d)} The adiabaticity index $\Gamma$ plotted versus $e_{\rm max}$ for $\tau = 10^4\,T_b$.}

\end{figure}
\clearpage
\end{document}